\documentclass[aip,pop,sd,reprint,10pt,amsmath,amssymb,graphicx]{revtex4-1}

\usepackage{bm}
\usepackage{dcolumn}
\usepackage{graphicx}
\usepackage{color}

\usepackage{mathtools}
\usepackage{amssymb}

\newcommand{\nc}{\newcommand}
\nc{\rcite}[1]{Ref.~\onlinecite{#1}}
\nc{\rcites}[1]{Refs.~\onlinecite{#1}}
\nc{\eqeqref}[1]{Eq.~\eqref{eq:#1}}
\nc{\eqseqref}[2]{Eqs.~\eqref{eq:#1}-\eqref{eq:#2} }
\nc{\secref}[1]{Sec.~\ref{sec:#1}}
\nc{\secsref}[2]{Sec.~\ref{sec:#1}-Sec.~\ref{sec:#2}}
\nc{\ssecref}[1]{Sec.~\ref{ssec:#1}}
\nc{\ssecsref}[2]{Sec.~\ref{ssec:#1}-Sec.~\ref{ssec:#2}}

\begin{document}

\title{Action Principles for Extended MHD Models}
\author{I.Keramidas Charidakos, M.Lingam, P.J.Morrison, R.L.White}
\affiliation{Institute for Fusion Studies and Department of Physics, The University of Texas at Austin, Austin, TX 78712, USA}
\author{A. Wurm}
\affiliation{Department of Physical and Biological Sciences, Western New England University, Springfield, MA 01119, USA}


\begin{abstract}
The general, non-dissipative, two-fluid model in plasma physics is Hamiltonian, but this property is sometimes lost or obscured in the process of deriving simplified (or reduced) two-fluid or one-fluid models from the two-fluid equations of motion. To ensure that the reduced models are Hamiltonian, we start with the general two-fluid action functional, and make all the approximations, changes of variables, and expansions directly within the action context. The resulting equations are then mapped to the Eulerian fluid variables using a novel nonlocal Lagrange-Euler map. Using this method, we recover L\"{u}st's general two-fluid model, extended MHD, Hall MHD, and electron MHD from a unified framework. The variational formulation allows us to use Noether's theorem to derive conserved quantities for each symmetry of the action.
\end{abstract}


\maketitle

\section{Introduction}

Fluid models are ubiquitous in the study of plasmas. It is desirable that the non-dissipative limits of fluid models be Hamiltonian, but this property is often lost in the process of deriving them (see e.g.\ Ref.~\onlinecite{kimura,pjmTTC14}). One way to ensure the Hamiltonian property of such models is to derive them from action principles, i.e, to start from a Hamiltonian parent-model action and to make all the approximations and manipulations directly in the action (see e.g., \rcites{morrison1998,morrison2005,morrison2009,morrison2014hamiltonian}). The equations of motion are then obtained as the stationary points of the action under variation with respect to the dynamical variables. This is known as {\it Hamilton's principle} in particle mechanics.

 Deriving the equations of motion of fluids\cite{taub1949,herivel1955,penfield1966} and plasmas\cite{low1958,sturrock1958,butcher1953,dougherty1974,newcomb1962,penfieldhaus1966,ilgisonis1999,morrison2009,morrison2005} from an action principle has a rich history. The reasons that such a formulation is pursued, even after the equations of motion are already known, are numerous. Finding conservation laws using Noether's theorem,\cite{padhye1996,padhye1996fluid,webb2007,webbII2005}, obtaining variational principles for equilibria\cite{andreussi2010,andreussi2012,morrison2014hamiltonian},  performing  stability  analyses\cite{elsasser1994,morrison2005,andreussi2012,kawazura,andreussi2013,squire2013,moawad}, or imposing constraints on a theory is straight-forward in the action functional, but often not easily done directly in the equations of motion.

 Fluids can be described within the Eulerian or the Lagrangian viewpoints. The former describes the fluid in terms of, e.g., the evolution of the fluid velocity field at a position $x$, whereas the latter tracks the motion of individual fluid elements. The map connecting these two viewpoints is known as the Lagrange-Euler map. Action functionals are naturally expressed in terms of Lagrangian variables, whereas the equations of motion of fluid models are Eulerian.

Here we are interested in fluid models describing two charged fluids, e.g., an ion and an electron fluid, interacting with an electromagnetic field. The general, non-dissipative two-fluid system is Hamiltonian, and therefore it is desirable that any model attempting to give a reduced description of it, should also be Hamiltonian and, consequently, not possess  spurious forms of dissipation. 

We will use the Lagrangian viewpoint and construct a general two-fluid action functional. Any subsequent ordering, approximation, and change of variables will be done directly in the action before deriving the equations of motion using Hamilton's principle.  To ensure that the final equations of motion are Eulerian, we only construct actions that can be completely expressed in terms of Eulerian variables. This general requirement was elucidated in 
Refs.~\onlinecite{morrison2009,morrison2014hamiltonian}, where it was termed the  {\it Eulerian Closure Principle}.

This paper is organized as follows: In \secref{twofluid} we review the Lagrangian and Eulerian picture of fluid mechanics, including the derivation of the two-fluid equations of motion through Lagrangian variations of a two-fluid action functional. Starting from this action, we derive a new one-fluid action functional using careful approximations, e.g., imposing quasineutrality, and a change of variables in \secref{newaction}. Here we also introduce a new Lagrange-Euler map and impose locality in order to derive Eulerian equations of motion in the new variables. In \secref{models} we show in detail how to derive various fluid models, e.g., extended MHD and Hall MHD, from this new one-fluid action functional. \secref{noether} contains a discussion of Noether's theorem applied to the new action functional, and finally, \secref{concl} our conclusions and some discussion of future work.

\section{Review: Two-fluid model and action}
\label{sec:twofluid}

In this section, we will briefly review the derivation of the non-dissipative two-fluid model equations of motion from the general two-fluid action functional. This action will be the starting point for deriving reduced models further below. In this context we establish our notation and, later on, discuss differences with our new procedure and results.   

Non-dissipative fluids can be described in two equivalent ways: The Eulerian (or spatial) point of view, which uses the physical observables of, e.g., fluid velocity $v(x,t)$ and mass density $\rho(x,t)$ as its fundamental variables and describes the fluid at an observation point $x$ in the three-dimensional domain as time passes, or the Lagrangian (or material) point of view, which considers individual fluid elements with position $q$ and tracks their time evolution. As described below, both pictures are related through the standard Lagrange-Euler map.

From an action functional/variational point of view, the Lagrangian picture is the more natural one, as it represents the infinite-dimensional generalization of the finite-dimensional Lagrangians of particle mechanics. The equations of motion are then obtained using Hamilton's principle as the stationary points of the action, i.e., the first variation of the action with respect to the variables is equal to zero.

We will use the Lagrangian picture as our starting point and construct a general two-fluid action functional. To ensure physical relevance of the
theory, we only construct actions that obey the Eulerian Closure Principle, which states that any action functional of a physical fluid theory must be completely expressible in Eulerian variables after the application of the Lagrange-Euler map.

To simplify our notation (consistent with \rcite{morrison1998}), we will avoid explicit vector notation and define the following:
$q_s = q_s(a,t)$ is the position of a fluid element ($s=(i,e)$ is the species label) in a rectangular coordinate system
where $a=\left(a_1,a_2,a_3\right)$ is any label identifying the fluid element and $q_s=\left(q_{s1},q_{s2},q_{s3}\right)$. Here we choose $a$ to be the
initial position of the fluid particle at $t=0$, although other choices are possible\cite{andreussi2013}. The Lagrangian velocity will then be denoted by $\dot{q}_s$.

The Eulerian velocity field will be denoted by $v_s(x,t)$ with $v_s=\left(v_{s1}, v_{s2}, v_{s3}\right)$ where $x=\left(x_1,x_2,x_3\right)$ is the position
in the Eulerian picture. Similarly, we define the electric field vector $E(x,t)$, the magnetic field vector $B(x,t)$, and the vector potential $A(x,t)$. If we need to explicitly refer to components of these vectors, we will use subscripts (or superscripts) $j$ and $k$. To simplify the equations, we will also often suppress the dependence on $x$, $a$, and $t$.

The action functional described below will include integrations over position space $\int d^3x$ and label space $\int d^3a$. We will not explicitly specify the domains of integration, but assume that our functions are well-defined on their respective domains, and that integrating them and taking functional derivatives is allowed. In addition, we will assume that all variations on the boundaries of the domains and any surface terms (due to integration by parts) vanish.

\subsection{Constructing the two-fluid action}

The action functional of a general theory of a charged fluid interacting with an electromagnetic field should
 include the following components: The energy of the electromagnetic field, the fluid-field interaction energy,
 the kinetic energy of the fluid, and the internal energy of the fluid, which describes the fluid's thermodynamic properties.

We will assume two independent fluids corresponding to two different species (ions and electrons with charge $e_s$, mass $m_s$ and
initial number density of $n_{s0}(a)$) which
interact with the electromagnetic field, but not directly with each other. Therefore the fluid-dependent parts of the action will naturally split
into two parts, one for each species.

The complete action functional is given by
\begin{equation}
\it{S}[q_s,A,\phi] = \int_T dt\; \it{L}\,,
\label{eq:gtfaction}
\end{equation}
where $T$ is a finite time interval and the Lagrangian $L$ is given by
\begin{align}
\it{L} &= \frac{1}{8\pi} \int d^3x\; \bigg[ \left| -\frac{1}{c}
\frac{\partial A(x,t)}{\partial t} -
\nabla \phi(x,t)\right|^2 
 \nonumber\\ 
&  \hspace{4cm}
-\left|\nabla\times A(x,t)\right|^2\bigg]
\label{field}
\\
\nonumber 
&+ \sum_s \int d^3a\:  n_{s0}(a)\int d^3x\; \delta\left(x-q_s(a,t)
\right) \\ 
&  \hspace{2.5cm} \times \left[
\frac{e_s}{c}\; \dot{q}_s\cdot A(x,t) - e_s \phi(x,t)
\right]
\label{coupling}
\\[.1in]
&+ \sum_s \int d^3a\;  n_{s0}(a) \Big[ \frac{m_s}{2}\;|{\dot{q}_s}|^2
\nonumber\\
&\hspace{2.5cm} 
-m_s U_s\left(m_s n_{s0}(a)/{\mathcal J}_s,s_{s0}\right)\Big]\,.
 \label{eq:gtfLag}
\end{align}

The symbol ${\mathcal J}_s$ is the Jacobian of the map between Lagrangian positions and labels, $q(a,t)$, which we will discuss in more detail below. Here we have expressed the electric and magnetic fields in terms of the vector and scalar potential, $E = -1/c \left(\partial A/\partial t\right) -\nabla \phi$ and $B = \nabla\times A$. The first term \eqref{field} 
 is the electromagnetic field energy, while the next expression \eqref{coupling} is the coupling of the fluid to the electromagnetic field,  which is achieved here by using the delta function. The last line of the Lagrangian $\it{L}$ \eqref{eq:gtfLag} represents the kinetic and internal energies  of the fluid.  Note, the specific internal energy (energy per unit mass) of species $s$, $U_s$, depends on the Eulerian density as well as a function $s_{s0}$, an entropy label for each species.   
Also note, that the vector and scalar potentials are Eulerian variables (i.e., functions of $x$). 

\subsection{Lagrange-Euler map}
\label{ssec:stLEmap}

In accordance with the above-mentioned Eulerian Closure Principle, we need to ensure that the action \eqseqref{gtfaction}{gtfLag}  can be completely expressed in terms of the desired set of Eulerian variables, which in turn ensures that the resulting equations of motion will also be completely Eulerian, hence representing a physically meaningful model.

The connection between the Lagrangian and Eulerian pictures of fluids is  the {\it Lagrange-Euler map}.  Before looking at the mathematical implementation of this map, it is instructive to discuss its meaning. As an example, consider the Eulerian velocity field $v(x,t)$ at a particular position $x$ at time $t$. The velocity of the fluid at that point will be the velocity of the particular fluid element $\dot{q}(a,t)$ which has started out at position $a$ at time $t=0$ and then arrived at point $x=q(a,t)$ at time $t$.

To implement this idea, we define the Eulerian number density $n_s(x,t)$ in terms of Lagrangian quantities as follows:
\begin{equation}
\label{eq:eulnumbdensI}
 n_s(x,t) = \int d^3a \;  n_{s0}(a) \; \delta\left(x-q_s(a,t)\right)\,.
\end{equation}
Using properties of the delta function, this relation can be rewritten as 
\begin{equation}
\label{eq:eulnumbdensII}
 n_s(x,t) = \left.\frac{n_{s0}(a)}{\mathcal{J}_s}\right|_{a = q^{-1}_s(x,t)}\,, 
\end{equation}
where, $\mathcal{J}_{s}=\det \left(\partial q_s /\partial a\right)$ is the Jacobian determinant.  Note that \eqeqref{eulnumbdensII} implies the continuity equation
\begin{equation}
\frac{\partial n_s}{\partial t} + \nabla \cdot \left(n_s v_s\right) = 0\,,
\end{equation}
which corresponds to local mass conservation if we define the mass density as $\rho_s=m_s n_s$.

The corresponding relation for the Eulerian velocity is 
\begin{equation}
\label{eq:LEvel}
 v_s(x,t)= \dot{q}_s(a,t)|_{a = q^{-1}_s(x,t)}\,, 
\end{equation}
where the dot means differentiation with respect to time at fixed particle label $a$. This relation follows from integrating out the delta function in
the definition of the Eulerian momentum density, $M_s:= m_s n_s v_s$,
\begin{equation}
 M_s(x,t) =  \int \!d^3a\, n_{s0}(a,t)   \delta\left(x-q_s(a,t)\right) m_s  \dot{q}_s(a,t)\,.
\end{equation}

Finally, our Eulerian entropy per unit mass, $s_s(x,t)$,  is defined by 
\begin{equation}
\label{eq:eulentrop}
\rho_s s_s(x,t) = \int\! d^3a \,  n_{s0}(a) s_{s0}(a) m_s \; \delta \left(x-q_s(a,t)\right)\,, 
\end{equation}
completing  our set of fluid Eulerian variables for this theory, which is $\{n_s,s_s, M_s\}$.  It is easy to check that the closure principle is satisfied by these variables. 

For later use, we quote (without proof) some results involving the determinant and its derivative 
\begin{equation}
\label{eq:cofactor}
 \frac{\partial q^k}{\partial a^j}\frac{A^i_{k}}{\mathcal{J}} = \delta^i_{j}\,,
\end{equation}
where $A^i_{k}$ is the cofactor of ${\partial q^k}/{\partial a^i}:=q^k_{,i}$. A convenient expression for $A^i_{k}$ is
\begin{equation}
 A^i_{k} = \frac{1}{2}\epsilon_{kjl}\epsilon^{imn}\frac{\partial q^j}{\partial a^m}\frac{\partial q^l}{\partial a^n}\,,
\end{equation}
where $\epsilon_{ijk} (= \epsilon^{ijk})$ is the Levi-Civita tensor. Using \eqeqref{cofactor} one can show that
\begin{equation}
\label{eq:derideterm}
 \frac{\partial \mathcal{J}}{\partial q^i_{,k}} = A^j_{i}
\end{equation}
and using the chain rule
\begin{equation}
\label{eq:derivq}
 \frac{\partial}{\partial q^k} = \frac{1}{\mathcal{J}}A^i_{k} \frac{\partial}{\partial a^i}\, .
\end{equation}
For further discussions see, e.g., \rcites{morrison1998,morrison2009,salmon1988}.

\subsection{Varying the two-fluid action}

The action of  \eqeqref{gtfaction} depends on four dynamical variables: the scalar and vector potentials, $\phi$ and $A$, and
the positions of the fluid elements $q_s$.

Varying with respect to $\phi$ yields Gauss's law
\begin{align*}
\partial_k\left(-\frac{1}{c}\frac{\partial A_k}{\partial t}-\partial_k\phi\right) & =
4\pi e\int d^3a\;  n_{i0}(a)\; \delta\left(x-q_i\right)\\
&- 4\pi e\int d^3a\;  n_{e0}(a) \; \delta\left(x-q_e\right)\,,
\end{align*}
where $\partial_k := \partial/\partial x^k$, or in more familiar form
\begin{equation}
\nabla\cdot E = 4\pi e \left(n_i(x,t)-n_e(x,t)\right)\, .
\end{equation}
Similarly, the variation with respect to $A$ recovers the Maxwell-Ampere law
\begin{align*}
\frac{1}{4\pi} &\left[ - \nabla\times\nabla\times A +\frac{1}{c}\frac{\partial}{\partial t}\left(-\frac{1}{c}\frac{\partial A}{\partial t}-
\nabla\phi\right)\right]\\
&- \frac{e}{c}\int d^3a\;n_{i0}\; \left[\delta\left(x-q_e\right) n_{e0} \dot{q}_e +
\delta\left(x-q_i\right)  \dot{q}_i\right] = 0
\end{align*}
or in more familiar form
\begin{equation}
\label{eq:amplaw}
\nabla\times B = \frac{4\pi J}{c} +  \frac{1}{c}\frac{\partial E}{\partial t}
\end{equation}
where the Eulerian current density $J$ is defined as
\begin{equation}
J(x,t) = e \left(n_i v_i-n_e v_e\right)\, .
\end{equation}
Recall that the other two Maxwell  equations are contained in the definition of the potentials.

Variation with respect to the $q_s$'s is slightly more complex, and we will show a few intermediate
steps. Varying the kinetic energy term is straight forward and yields
\begin{equation}
\label{eq:tfkinetic}
-n_{s0}(a)m_s\ddot{q}_s(a,t)
\end{equation}
The $j$-th component of the interaction term results in
\begin{align}
& e_s n_{s0}(a) \left[ -\frac{1}{c}\frac{dA^j(q_s,t)}{dt} +
 \frac{1}{c}\dot{q}_s^k\frac{\partial A^k(q_s,t)}{\partial q_s^j} - \frac{\partial \phi(q_s,t)}{\partial q_s^j} \right]
 \nonumber\\
& \quad = e_s n_{s0}(a) \left[-\frac{1}{c}\frac{\partial A^j(q_s,t)}{\partial t} - \frac{1}{c}\dot{q}^k_s\frac{\partial A^j(q_s,t)}{\partial q_s^k}\right.
\nonumber\\
& \qquad\qquad\qquad\qquad  \left.+\frac{1}{c}\dot{q}_s^k\frac{\partial A^k(q_s,t)}{\partial q_s^j} - \frac{\partial \phi(q_s,t)}{\partial q_s^j}\right]
\label{eq:interaction}\\
& \quad = e_s n_{s0}(a) \left[ E(q_s,t) + \frac{1}{c} \dot{q}_s(a,t)\times\left(\nabla_{q_s}\times A\left(q_s,t\right)\right)\right]_j
\nonumber
\end{align}
Note that this expression is purely Lagrangian. The fields $A$ and $E$ are evaluated at the positions $q_s$ of the fluid elements and the curl $\nabla_{q_s}\times$ is taken with respect to the Lagrangian position. Also note, since $q_s=q_s(a,t)$, any total time derivative of, e.g., $A(q_s,t)$ will
result in two terms.

Variation of the internal energy term yields 
\begin{equation}
\label{eq:intenergy}
 A^j_{i}\frac{\partial}{\partial a_j}\left(\frac{\rho_{s0}^2}{\mathcal{J}_s^2}
\frac{\partial U\left(\frac{\rho_{s0}}{\mathcal{J}_s},s_{s0}\right)}{\partial \left(\frac{\rho_{s0}}{\mathcal{J}_s}\right)}\right)\, .
\end{equation}

Setting the sum of \eqseqref{tfkinetic}{intenergy} equal to zero and invoking the usual thermodynamic relations
between internal energy and pressure and temperature, 
\begin{equation}
\label{eq:pressure}
 p_s = (m_s n_s)^2 \frac{\partial U_s}{\partial (m_s n_s)}
 \quad\mathrm{and}\quad
 T_s=\frac{\partial U_s}{\partial s_s}
\end{equation}
results in the well-known (non-dissipative) two-fluid equations of motion
\begin{equation}
m_{s}n_{s}\left(\frac{\partial v_s}{\partial t}+ v_s\cdot\nabla v_s\right)=e_{s}n_{s}\left( E+\frac{1}{c} v_s
\times B\right)-\nabla p_{s}
\end{equation}

Further analysis (see e.g.\ \rcites{lust1959,goedbloed2004}) of these equations usually involves the addition and subtraction of the two-fluid equations and a change of variable transformation to
\begin{align}
V & = \frac{1}{\rho_m}\;\left( m_i n_i v_i + m_e n_e v_e\right)\nonumber\\
J & =e\left(n_i v_i-n_e v_e\right)\nonumber\\
\rho_m & = m_i n_i +m_e n_e \label{eq:euchangofvar}\\
\rho_q & = e \left(n_i-n_e\right)\,.\nonumber
\end{align}

The resulting equations are then simplified by, e.g., making certain assumptions (quasineutrality,
$v<<c$, etc.) and ordering to obtain two new {\it one}-fluid equations --  one often referred to as
the {\it one-fluid momentum equation} and the other as the {\it generalized Ohm's law}.

\section{The new one-fluid action}
\label{sec:newaction}

The first step in building an action functional for fluid models is to decide on the relevant Eulerian observables of the model. Since we want to  derive, e.g., the two-fluid model of L\"{u}st and various reductions, our Eulerian observables are going to be the set $\{n,s,s_e,V, J, E, B\}$, where $s=(m_is_i+s_em_e)/m$, with $m=m_e +m_i$,  and $n$ is a single number density variable. 

Next we have to define our Lagrangian variables and with them construct the action. Any additional assumption (e.g., quasineutrality etc.) and ordering will be implemented on the action level. Varying the new action will then result in equations of motion that, using properly defined Lagrange-Euler maps, will Eulerianize to, e.g.,  L{\"u}st's equation of motion and the generalized Ohm's law.

\subsection{New Lagrangian variables}
\label{ssec:newlagvar}

We will start by defining a new set of Lagrangian variables, inspired by \eqeqref{euchangofvar}\footnote{First presented in A. Wurm and P.J. Morrison, Action principle derivation of one-fluid models from two-fluid actions, Bulletin of the Am. Phys. Soc, Vol. 54, Nr. 4 (2009).}, 
\begin{align}
 Q(a,t) & =  \frac{1}{\rho_{m0}(a)} \left( m_i n_{i0}(a) q_i(a,t) +
m_e n_{e0}(a)q_e(a,t)\right)\nonumber\\[.12in]
D(a,t) & =  e
\left(n_{i0}(a) q_i(a,t)-n_{e0}(a) q_e(a,t)\right)\nonumber\\[.12in]
\rho_{m0}(a) & =  m_i n_{i0}(a)+m_e n_{e0}(a)\label{eq:newfullvar}\\[.12in]
\rho_{q0}(a) & =  e \left(n_{i0}(a)-n_{e0}(a)\right)\,.\nonumber
\end{align}
Here $Q(a,t)$ can be interpreted as a center of mass position variable and $D(a,t)$ as a local dipole moment variable, connecting an ion fluid
element to an electron fluid element. It is then straight-forward to take the time-derivative of $Q$ and $D$ which can be interpreted
as the center-of-mass velocity $\dot{Q}(a,t)$ and a Lagrangian current $\dot{D}(a,t)$, respectively. Using appropriately defined Lagrange-Euler
maps, $\dot{Q}(a,t)$ will map to the Eulerian velocity $V(x,t)$ and $\dot{D}(a,t)$ to the Eulerian current $J(x,t)$ as defined by \eqeqref{euchangofvar}.

We will also need the inverse of this transformation, 
\begin{align}
q_i(a,t)& = \frac{\rho_{m0}(a) Q(a,t) + \frac{m_e}{e} D(a,t)}{\rho_{m0}(a)+ \frac{m_e}{e} \rho_{q0}}\nonumber\\[.12in]
q_e(a,t) & = \frac{\rho_{m0}(a) Q(a,t) - \frac{m_i}{e} D(a,t)}{\rho_{m0}(a)- \frac{m_i}{e} \rho_{q0}}\nonumber\\[.12in]
n_{i0}(a) & = \frac{\rho_{m0}(a)+ \frac{m_e}{e}\rho_{q0}(a)}{m}\label{eq:invvar}\\[.12in]
n_{e0}(a)& = \frac{\rho_{m0}(a)- \frac{m_e}{e}\rho_{q0}(a)}{m}\nonumber\,.
\end{align}

\subsection{Ordering of fields and quasineutrality}

Typically, reductions of the full two-fluid model are obtained by imposing an auxiliary ordering on the equations of motion. In order to preserve the variational formulation, we perform an ordering directly in the action.

To construct the action, we will start with the two-fluid action  of \eqeqref{gtfaction} and change variables to $Q$ and $D$, but in light  of the fluid models we are interested in, we will first make two simplifying assumptions: We  order the fields in the action so  that the displacement current in \eqeqref{amplaw} will vanish, and we  assume quasineutrality. In this section, we  describe this field ordering in detail and  discuss quasineutrality in the Lagrangian variable context, which as far as we know  has not  been done before.

The omission of the displacement current is allowed, when the time scale of changes in the field configuration is long relative to the time it takes for radiation to ``communicate" these changes across the system\cite{roberts1967}. We  use non-dimensional variables by introducing a characteristic scale $B_0$ for the magnetic field and a characteristic length scale $\ell$ for gradients. Times are then normalized by the Alfv\'en time $t_A = {B_0}/{\sqrt{4\pi \rho}}$ and the $\dot{q}_s$'s by the Alfv\'en speed $v_A =  {\ell}/{t_A}$, resulting in the following form for the sum of the field and interaction terms of the Lagrangian \eqref{eq:gtfLag}:
\begin{align*}
& \frac{B_0^2}{8\pi}\int dt \int d^3\hat{x}\left[\left|-\frac{v_A}{c}\frac{\partial\hat{A}}{\partial \hat{t}}-\frac{\phi_0}{B_0 \ell}\hat{\nabla}\hat{\phi}\right|^2 
- \left|\hat{\nabla}\times{\hat{A}}\right|^2\right]\\
& \hspace{.5cm} +\sum_s B_0^2\left[\int dt \int d^3\hat{a}\;n_0\hat{n}_{s0}(a)e_{s}\int d^3\hat{x}\;\delta\left({\hat{x}}-{\hat{q}}_{s}\right)\right.\\
&  \hspace{1.5cm}\left.\times\left(\frac{v_A \ell}{B_0 c}{\hat{\dot{q}}}_{s}\cdot{\hat{A}}-\frac{\phi_o}{B_0^2 }\hat{\phi}\right)\right]\,, 
\end{align*}
where $\phi_0$ and $n_0$ are yet to be specified scales for the electrostatic potential and the densities of both species, respectively. We also require that the two species' velocities are of the same scale.  
Requiring the two interaction terms in the Lagrangian to be of the same order  results in a scaling for $\phi$; viz.,  $\phi_0 \equiv B_0 \ell  {v_A}/{c}$. Thus,  both parts of the $|E|^2$ term are of order $\mathcal{O}\left({v_A}/{c}\right)$. Neglecting this term and varying with respect to $\hat{A}$ results in 
\begin{align*}
\hat{\nabla}\times\hat{B} & = \frac{4\pi e n_0 v_A}{c}\frac{\ell}{B_0}\left(\int d^3a\; \delta(\hat{x}
- \hat{q_i})\hat{n}_{i0}(a)\hat{\dot{q_i}}\right.\\
& \hspace{ 1.25cm} \left. - \int d^3a \delta(\hat{x}- \hat{q_e})\hat{n}_{e0}(a)\hat{\dot{q_e}}\right)\,,
\end{align*}
which can be written as
\begin{equation}
\label{eq:nablaJ}
\frac{B_0}{\ell}\hat{\nabla}\times\hat{B} = \frac{4\pi j_0}{c} \hat{J}\,,
\end{equation}
where $j_0 = e n_0 v_A$ is a scale for the current.


Varying the scaled action with respect to $\hat{\phi}$ yields
\begin{equation}
0 =   \int d^3\hat{a} \delta(\hat{x}- \hat{q_i}) \hat{n}_i - \int d^3\hat{a} \delta(\hat{x}- \hat{q_e})\hat{n}_e  \equiv \Delta \hat{n}
\end{equation}
The above equation states that the difference in the two densities is zero, i.e., the plasma is quasineutral, 
 a property that holds locally, i.e., $n_i(x,t)=n_e(x,t)$. Using \eqeqref{eulnumbdensII}, this statement would correspond to the following in the Lagrangian variable picture:
\begin{equation}
\left.\frac{n_{i0}(a)}{\mathcal{J}_i(a,t)}\right|_{a=q_i^{-1}(x,t)} 
= \left.\frac{n_{e0}(a)}{\mathcal{J}_e(a,t)}\right|_{a=q_e^{-1}(x,t)}\,.
\label{eq:lagquasin}
\end{equation}

In the Lagrangian picture we will make the additional assumption of homogeneity: 
$n_{i0}(a) = n_{e0}(a) =$ constant, which is  natural  for  the plasma  we are modeling.  It states that at $t=0$ all fluid elements are identical in the amount of density  they carry.   Therefore, $n_{i0}$ and $n_{e0}$ can be replaced by a constant $n_0$. Equation \eqref{eq:lagquasin} then reduces to a statement about the two Jacobians
\begin{equation}
{\mathcal{J}_i(a,t)}\Big|_{a=q_i^{-1}(x,t)} =  {\mathcal{J}_e(a,t)}\Big|_{a=q_e^{-1}(x,t)}\,,
 \label{eq:jacobians}
\end{equation}
which will play a central role in our development  below.

Note, that the homogeneity assumption ($n_{i0} = n_{e0} = n_0$) does not prohibit us from describing quasineutral plasmas with density gradients. What we would have to do in this case, would be to pick our labeling scheme, and hence the Jacobian, accordingly, as to reflect the initial density gradient of the configuration.  Thus there is freedom in this regard beyond what we are assuming now.

\subsection{Action functional}

We are now ready to implement the change of variables discussed in \ssecref{newlagvar}.
Because of the homogeneity assumption $n_{i0}(a) = n_{e0}(a) = n_0$, the new variables 
of  \eqeqref{newfullvar} reduce to
\begin{align}
Q(a,t) &= \frac{m_i}{m} q_i(a,t)+\frac{m_e}{m} q_e(a,t)\nonumber\\
D(a,t) &= e n_0\left(q_i(a,t)- q_e(a,t)\right)\nonumber\\
\rho_{m0}(a) &=m n_0\label{eq:newvarhom}\\
\rho_{q0}(a) &=0\nonumber
\end{align}
and the inverse transformation of  \eqeqref{invvar} to
\begin{align}
q_i(Q,D):= q_i(a,t) &= Q(a,t)+ \frac{m_e}{men_0}D(a,t)\nonumber\\
 q_e(Q,D):=q_e(a,t) &= Q(a,t) -\frac{m_i}{men_0}D(a,t)\,,
 \label{eq:invarhom}
\end{align}
where we choose the notation $q_s(Q,D)$ to emphasize that the $q_s$
should not be thought of as ion/electron trajectories any more but as specific
linear combinations of $Q(a,t)$ and $D(a,t)$. In addition, we will need the ion and electron Jacobians, $\mathcal{J}_i(Q,D)$ and $\mathcal{J}_e(Q,D)$, now expressed in terms of $Q$ and $D$.

The resulting action functional has the form:
\begin{widetext}
\begin{align}
 S = & -\frac{1}{8\pi} \int dt \int d^3x\; \left|
\nabla\times A(x,t)\right|^2\nonumber\\[.1in]
& + \int dt \int d^3x \int d^3a  \; n_{0} \left\{\delta\!\left(x-q_i(Q,D)\right)
                     \left[\frac{e}{c}\left(\dot{Q}(a,t)+\frac{m_e}{m e n_0}\dot{D}(a,t)\right)\!\cdot\!
                         A(x,t)-e\phi(x,t)\right]\right\}\nonumber\\[.1in]
& +  \int dt \int d^3x \int d^3a  \; n_{0} \left\{\delta\!\left(x-q_e(Q,D)\right)
                    \left[-\frac{e}{c}\left(\dot{Q}(a,t)-\frac{m_i}{m e n_0}\dot{D}(a,t)\right)\!\cdot\!
                                    A(x,t)+e\phi(x,t)\!\right]\right\}\nonumber\\[.1in]
 & \hspace{1cm}+\frac{1}{2} \int dt \int d^3a\;n_0 \left[ m_i|\dot{Q}|^2(a,t)+
      \frac{m_i m_e}{m e^2 n^2_0}|\dot{D}|^2(a,t)\right]\nonumber\\[.1in]
 & \hspace{1.5cm} -\int dt \int d^3a\;n_{0}  \left[ m_i U_{i}\left(\frac{m_i n_{0}}{\mathcal{J}_i(Q,D)},(ms_0-m_es_{e0})/m_i\right)
 +m_e U_{e}\left(\frac{m_e n_{0}}{\mathcal{J}_e(Q,D)},s_{e0}\right)\right]\,,
 \label{eq:newgenact}
\end{align}
\end{widetext}
 where recall $s_0=(m_is_{i0}+ m_es_{e0})/m$.

\subsection{Nonlocal Lagrange-Euler maps}

 Now we  define the Lagrange-Euler maps that connect the Eulerian observables $V$ and $J$ to the new Lagrangian variables $Q$ and $D$. Referring to \ssecref{stLEmap}, one can see that a Lagrange-Euler map is a relationship between a Lagrangian quantity and some Eulerian observables,  which holds only when it is evaluated on a trajectory $x=q_s(a,t)$.
 If we apply the inverse Lagrange-Euler maps from Eqs.~(\ref{eq:eulnumbdensII}) and (\ref{eq:LEvel}) to \eqeqref{euchangofvar} and assume quasineutrality, we get
\begin{align}
V(x,t) &= \left.\frac{m_i}{m}\, \dot{q}_i(a,t)\right|_{a=q^{-1}_i(x,t)} +
\left.\frac{m_e}{m} \, \dot{q}_e(a,t)\right|_{a=q^{-1}_e(x,t)}\nonumber\\
J(x,t) & = \left.e\left(\frac{n_0}{\mathcal{J}_i(a,t)}\dot{q}_i(a,t)\right)\right|_{a=q^{-1}_i(x,t)}\nonumber \\
& \hspace{2 cm}\left. -e\left(\frac{n_0}{\mathcal{J}_e(a,t)} \dot{q}_e(a,t)\right)\right|_{a=q^{-1}_e(x,t)}\nonumber\\
n(x,t) &= \left.\frac{m_i}{m}\left(\frac{n_0}{\mathcal{J}_i(a,t)}\right)\right|_{a=q^{-1}_i(x,t)} \label{eq:numdensity}\\
 & \hspace{2.72 cm} \left.+ \frac{m_e}{m}\left(\frac{n_0}{\mathcal{J}_e(a,t)}\right)\right|_{a=q^{-1}_e(x,t)}
\nonumber\\
s(x,t) &= \frac{m_i }{m}\, s_{i0}\Big{|}_{a=q^{-1}_i(x,t)} + \frac{m_e }{m}\, s_{e0}\Big{|}_{a=q^{-1}_e(x,t)} 
 \label{eq:ents}\\
s_e(x,t) &=  s_{e0} \Big|_{a=q^{-1}_e(x,t)} \,.
 \label{eq:entse}
\end{align}

 The definitions of $Q(a,t)$ and $D(a,t)$ in \eqeqref{newvarhom} suggest that their time-derivatives should be associated with $V$ and $J$, respectively.
 However, both $\dot{Q}$ and $\dot{D}$ are nonlocal objects, since they relate the velocities of electrons and ions which
 are located at different points in space. This means that neither $\dot{Q}$, nor $\dot{D}$, when evaluated at the inverse maps for $a$, can
Eulerianize to a local velocity or current, since, in general, $x = q_i(Q,D)$ and $x' = q_e(Q,D)$ with $x\neq x'$ or, they are simultaneously evaluated at different trajectories.
 Therefore, we have two different inverse functions where the Lagrangian quantities are to be evaluated, namely, $a = q_i^{-1}(x,t)$ and $a = q_e^{-1}(x',t)$ which should be thought of as the inverse functions of $x=q_i(Q,D)$ and $x'=q_e(Q,D)$.
To make this work, we {\it define} our Lagrange-Euler maps with $x=x'$ as
\begin{align}
V(x,t) &= \left.\frac{m_i}{m}\left( \dot{Q}(a,t) + \frac{m_e}{m e n_0}\dot{D}(a,t)\right)\right|_{a=q^{-1}_i(x,t)}
\label{eq:LEmap}\\
&  + \left.\frac{m_e}{m} \left(\dot{Q}(a,t) - \frac{m_i}{m e n_0}\dot{D}(a,t)\right)\right|_{a=q^{-1}_e(x,t)} \nonumber\\
J(x,t) &= \left. \frac{e n_0}{\mathcal{J}_i(a,t)}\left(\dot{Q}(a,t) + \frac{m_e}{m e n_0}\dot{D}(a,t)\right)\right|_{a=q^{-1}_i(x,t)}\nonumber\\
& - \left.
\frac{e n_0}{\mathcal{J}_e(a,t)}\left(\dot{Q}(a,t) - \frac{m_i}{m e n_0}\dot{D}(a,t)\right)\right|_{a=q^{-1}_e(x,t)}\,.
\nonumber
\end{align}
Due to \eqeqref{jacobians}, the two Jacobian determinants are equal  (as long as they are evaluated at the respective inverse functions) and can be replaced by a common Jacobian determinant, $\mathcal{J}$.

 The maps just defined are straight-forward to apply for mapping an Eulerian statement to a Lagrangian one,  but for our purpose, we have to invert them. To keep careful track of the two inverse functions, we first invert 
 the intermediate relations
\begin{align}
V(x,t) & + \frac{m_e}{m e n(x,t)} J(x,t)\nonumber\\
 &= \left.\left(\dot{Q}(a,t) + \frac{m_e}{m e n_0}\dot{D}(a,t)\right)\right|_{a = q_i^{-1}(x,t)}
 \label{eq:intmapI}
\\
V(x,t) & - \frac{m_i }{m e n(x,t)} J(x,t)\nonumber\\
 & =  \left.\left(\dot{Q}(a,t) - \frac{m_i}{m e n_0}\dot{D}(a,t)\right)\right|_{a = q_e^{-1}(x,t)}\,,
 \label{eq:intmapII}
\end{align}
where we have used \eqeqref{eulnumbdensII}. The inverse Lagrange-Euler maps are now given by
\begin{align}
 \dot{Q}(a,t) &= \left.\frac{m_i}{m}\left(V(x,t)+\frac{m_e}{men(x,t)}J(x,t)\right)\right|_{x=q_i(Q,D)}
 \nonumber\\
  & + \left.\frac{m_e}{m}\left(V(x',t)-\frac{m_i}{men(x',t)}J(x',t)\right)\right|_{x'=q_e(Q,D)}
  \nonumber\\
\dot{D}(a,t)&= \left.en_0\left(V(x,t)+\frac{m_e}{men(x,t)}J(x,t)\right)\right|_{x=q_i(Q,D)}
\nonumber\\
&- \left.en_0\left(V(x',t)-\frac{m_i}{men(x',t)}J(x',t)\right)\right|_{x'=q_e(Q,D)}
\,.
\label{eq:invLEmap}
\end{align}
Note that the construction of the maps  of Eqs.~\eqref{eq:LEmap} and \eqref{eq:invLEmap} can be done with any invertible linear combination of the time derivatives of our Lagrangian variables. The only restriction is that the action should comply with the Eulerian Closure Principle, i.e., it should be expressible entirely in terms of the Eulerian observables. It is straightforward to show that this is true in our case.

\subsection{Lagrange-Euler maps without quasineutrality}

Had we  not assumed quasineutrality, we would have to proceed differently: \eqeqref{euchangofvar} implies that the proper
Lagrangian variables that would Eulerianize to velocity and current would be 
\begin{align*}
V(x,t) &= \frac{\left.m_i\left(\frac{n_{i0}}{\mathcal{J}_i} \dot{q}_i(a,t)\right)\right|_{a=q^{-1}_i(x,t)}}
{\left.m_i\left(\frac{n_{i0}}{\mathcal{J}_i}\right)\right|_{a=q^{-1}_i(x,t)} +
\left.m_e\left(\frac{n_{e0}}{\mathcal{J}_e}\right)\right|_{a=q^{-1}_e(x,t)}}  \\
&+\frac{\left.m_e\left(\frac{n_{e0}}{\mathcal{J}_e} \dot{q}_e(a,t)\right)\right|_{a = q^{-1}_e(x,t)}}
{\left.m_i\left(\frac{n_{i0}}{\mathcal{J}_i}\right)\right|_{a=q^{-1}_i(x,t)} +
\left.m_e\left(\frac{n_{e0}}{\mathcal{J}_e}\right)\right|_{a=q^{-1}_e(x,t)}}
\,,\\
J(x,t) &=\left. e\left(\frac{n_{i0}}{\mathcal{J}_i} \dot{q}_i(a,t)\right)\right|_{a=q^{-1}_i(x,t)}\\
&\hspace{2.75cm} -
\left.e\left(\frac{n_{e0}}{\mathcal{J}_e} \dot{q}(a,t)\right)\right|_{a=q^{-1}_e(x,t)}\,.
 \end{align*}
The above equations suggest that without quasineutrality, the definitions for $\dot{Q}$, $\dot{D}$,  etc. should be modified to the following:
\begin{align*}
 \dot{Q}(a,t) &= \frac{1}{\rho_{m0} (a)}\big(m_i \mathcal{J}_e n_{i0}(a)\dot{q}_i(a,t)\\
 &\hspace{2cm}+ m_e \mathcal{J}_i n_{e0}(a)\dot{q}_e(a,t)\big)\\
\dot{D}(a,t) &= e\big( \mathcal{J}_e n_{i0}(a)\dot{q}_i(a,t)- \mathcal{J}_i n_{e0}(a)\dot{q}_e(a,t)\big)\\
\rho_{m0}(a) &= m_i \mathcal{J}_e n_{i0}(a)+m_e \mathcal{J}_i n_{e0}(a)
\end{align*}
where $\dot{Q}/\left(\mathcal{J}_i\mathcal{J}_e\right)$ maps to $V(x,t)$ and $\dot{D}/\left(\mathcal{J}_i\mathcal{J}_e\right)$ to $J(x,t)$. In this  case,  however, both $\dot{Q}$ and $\dot{D}$ are implicitly defined, since $\mathcal{J}_i$ and $\mathcal{J}_e$ depend on them. This problem is absent when only manipulating the Eulerian
equations of motion. It might suggest though that when quasineutrality does not hold, the one-fluid description might not be appropriate. This can also be seen in the most general case derived by L\"{u}st in \rcite{lust1959}. The resulting equations of motion in $V$ and $J$ still contain terms explicitly referring to ion/electron quantities, e.g., $n_i$ and $n_e$.  From a  variational point of view, it is not obvious how to apply the Eulerian Closure Principle without quasineutrality. It seems that in order to preserve it, one would need to distinguish between integrations over ion and electron labels, so that the
$d^3a$ could be related to the proper $\mathcal{J}_s$.

\subsection{Derivation of the continuity and entropy equations}

Before we derive the equations of motion for several different models in the next section, we derive here the continuity equation, which all of the models below have in common, and the entropy equations. 

Due to the identity of the Jacobians from \eqeqref{jacobians},  the equation for $n$ (\eqeqref{numdensity}) reduces to
 \begin{align}
n(x,t) &= \!\left.\left(\frac{n_0}{\mathcal{J}_i(a,t)}\right)\right|_{a=q^{-1}_i(x,t)}\!\! \!=\left. \left(\frac{n_0}{\mathcal{J}_e(a,t)}\right)\right|_{a=q^{-1}_e(x,t)}\,,
\label{eq:numdensityII}
\end{align}
where $q_s^{-1}$ are still the inverse functions of $q_s(Q,D)$. Inverting the equation for the ions and taking the time derivative yields
 \[
\left.\frac{dn}{dt}\right|_{x=q_i(Q,D)} = \frac{d}{dt}\frac{n_0}{\mathcal{J}_i(a,t)} = -\frac{n_0}{\mathcal{J}_i^2(a,t)}\frac{\partial\mathcal{J}_i}{\partial t}\, .
\]
 To Eulerianize the equation above, we use the well-known relations $d/dt = \partial/\partial t + v\cdot\nabla$ and $\partial \mathcal{J}/\partial t = \mathcal{J}\nabla\cdot v$. The key here is to use the correct Eulerian velocity, in this case the ion velocity in terms of $V$ and $J$. The result is
\[
\frac{\partial n}{\partial t} + \left(V+\frac{m_e}{m e n}J\right)\cdot \nabla n = -n\nabla\cdot\left(V+\frac{m_e}{m e n}J\right)
\]
which can be further reduced to
\[
\frac{\partial n}{\partial t} + \nabla\cdot(n V) + \frac{m_e}{m e}\nabla\cdot J = 0\, .
\]
However, we already know from \eqeqref{nablaJ} that $\nabla\cdot J = 0$. Therefore, no matter which equality we choose in \eqeqref{numdensityII}, the continuity equation will be the same, 
\begin{equation}
\label{eq:newconteq}
\frac{\partial n}{\partial t} + \nabla\cdot(n V) = 0
\end{equation}

Similarly, from \eqeqref{ents} we obtain 
\[
\frac{\partial s}{\partial t} + V\cdot \nabla s=0
\]
and from \eqeqref{entse}
\[
\frac{\partial s_e}{\partial t} + \left(V- \frac{m_i}{men}J\right) \cdot \nabla s_e=0\,,
\]
or to leading order in $m_e/m_i$
\[
\frac{\partial s_e}{\partial t} + \left(V- \frac{1}{en}J\right) \cdot \nabla s_e=0\,.
\]

\section{Derivation of reduced models}
\label{sec:models}

If we vary the action functional \eqref{eq:newgenact} with respect to $Q$ and $D$ and subsequently apply the Lagrange-Euler map we recover the momentum equation and generalized Ohm's law of L\"{u}st\footnote{Note,  there are typos in Eqs.~(2.9) and (2.10) of  Ref.~\onlinecite{lust1959}  that prevent the term $N_1$ from vanishing when imposing quasineutrality.} (in the non-dissipative limit):
\begin{eqnarray}
 && \hspace{ -2.4 cm} n m \left(\frac{\partial V}{\partial t} + (V\cdot\nabla) V\right)   
\\
&=&-\nabla p + \frac{J\times B}{c} - \frac{m_im_e}{me^2}(J\cdot\nabla)\left(\frac{J}{n}\right)
 \nonumber\\
E +\frac{V\times B}{c} &=&
\frac{m_i m_e}{m e^2 n}\left(\frac{\partial J}{\partial t} + (J\cdot\nabla)V - (J\cdot\nabla)\left(\frac{J}{n}\right)\right)
\nonumber \\
&& \hspace{ -.2 cm} + \frac{m_i m_e}{m e^2}(V\cdot\nabla)\left(\frac{J}{n}\right)
+ \frac{(m_i-m_e)}{m e n c} (J\times B)
\nonumber\\
&&  \hspace{1.2cm}- \frac{m_i}{men}\nabla p_e + \frac{m_e}{men}\nabla p_i\,.
\end{eqnarray}
We will not show this lengthy, although straightforward,  calculation here, but instead show the detailed derivation
of extended MHD in the next section which requires one more ordering in the action  of \eqeqref{newgenact}.

\subsection{Extended MHD}

At this point we will make one more simplification: We define the mass ratio $\mu = m_e/m_i$ and
order the action functional keeping terms up to first order in $\mu$. Up to first order, the change
of variables is
\begin{align}
 q_i(Q,D) &= Q(a,t) +\frac{\mu}{en_0}D(a,t)\nonumber\\
 q_e(Q,D) &= Q(a,t) -\frac{1-\mu}{en_0}D(a,t)\label{eq:qiqefirstorder}
\end{align}
and the action takes on the form
\begin{align}
 S &= -\frac{1}{8\pi} \int dt \int d^3x\; \left|
\nabla\times A(x,t)\right|^2\nonumber\\[.1in]
& + \int dt \int d^3x \int d^3a  \; n_{0}\, \biggl{\{}\delta\!\left(x-q_i(Q,D)\right)\nonumber\\
&\quad\left.\times\left[\frac{e}{c}\dot{Q}(a,t)+\frac{\mu}{c n_0}\dot{D}(a,t)\!\cdot\!
A(x,t)-e\phi(x,t)\right]\right\}\nonumber\\[.1in]
& +  \int dt\int d^3x\int d^3a  \; n_{0}\,  \biggl{\{}\delta\!\left(x-q_e(Q,D)\right)
\nonumber\\
&\quad \left.\times\left[-\frac{e}{c}\dot{Q}(a,t)+\frac{(1-\mu)}{c n_0}\dot{D}(a,t)\!\cdot\!
A(x,t)+e\phi(x,t)\!\right]\right\}\nonumber\\[.1in]
 & \!+\frac{1}{2} \int\! \!dt\!\int\! \!d^3a\, n_0  m_i\!\left(\!\!(1+\mu)|\dot{Q}|^2(a,t) + \!
      \frac{\mu}{e^2 n^2_0}|\dot{D}|^2(a,t)\right)
      \nonumber\\[.1in]
 & -\int \!\!dt\!\int\!\!d^3a \, n_0 \bigg(\mathfrak{U}_{e}\left(\frac{n_{0}}{\mathcal{J}_e(Q,D)},s_{e0}\right) 
 \nonumber\\
 &\hspace{3.5 cm}+  \mathfrak{U}_{i}\left(\frac{n_{0}}{\mathcal{J}_i(Q,D)},s_{i0}\right)\bigg)\,.
\label{eq:actionextmhd}
\end{align}
where for convenience we have replaced the $U_s$, the internal energy per unit mass, by $\mathfrak{U}_s$, the internal energy per  particle.  The pressure is obtained from the latter according to $p_s=n^2 \partial \mathfrak{U}_s/\partial n$. 

Varying the action with respect to $Q_k$ yields
\begin{align}
 0 &= -n_0 m_i(1+\mu)\ddot{Q}_k(a,t) -\partial_k  p \nonumber\\[.1in]
 & + n_0 \left[\frac{e}{c}\left(\dot{Q}_j(a,t)+\frac{\mu}{e n_0}\dot{D}_{j}(a,t)\right)
\frac{\partial A_j(x,t)}{\partial x^k}\right.\nonumber\\
& \qquad\qquad\left. - e\partial_k\phi(x,t) - \frac{e}{c}\frac{d}{dt}A_k(x,t)\right]\Biggl{|}_{x=q_i(Q,D)}
\nonumber\\[.1in]
 & + n_0 \left[-\frac{e}{c}\left(\dot{Q}_j(a,t)-\frac{(1-\mu)}{e n_0}\dot{D}_j(a,t)\right)
\frac{\partial A_j(x,t)}{\partial x^k}\right.\nonumber\\
& \qquad\qquad\left. + e\partial_k\phi(x,t) + \frac{e}{c}\frac{d}{dt}A_k(x,t)\right]\Biggl{|}_{x=q_e(Q,D)}\,.
\label{eq:qeqextmhd}
\end{align}

The variation of the internal energy term proceeds by varying $q_s$ through Eqs.~\eqref{eq:qiqefirstorder}, giving $\delta q_s=\delta Q$ and using these expressions in the variation of the Jacobians $\mathcal{J}_s$.   We have given the Eulerian result since the Lagrangian one has two terms of the form of  \eqeqref{intenergy},  and  it  is  cumbersome to carry this through the rest of the calculation. (See  Ref.~\onlinecite{kimura} for a treatment that orders  the Eulerian equations directly.)  Consistent with Dalton's law, the total single fluid pressure is $p=p_i+p_e$ and both these pressures  come in entirely at the zeroth  order of $\mu$.  Note, that the two time derivatives of $A$ do not cancel, because they are advected by different flows, or, since we are still in the Lagrangian framework, they are evaluated at different arguments.

To find the Eulerian equations of motion, we start with \eqeqref{invLEmap}  (up to first order in $\mu$) and impose locality, i.e., $x=x'$, such that $\dot{Q}$ maps to $V(x,t)$ and $\dot{D}$ to $J(x,t)$. However, the time derivatives of $\dot{Q}$ and $\dot{D}$ have to be treated with care as they each consist
of two terms that  are advected with different velocities. We will  show how to Eulerianize the equations of motion in detail.

The $\ddot{Q}$ in the first term of \eqeqref{qeqextmhd} can be re-written as
\begin{align}
&\ddot{Q}(a,t) \\
& =\left.\left(1-\mu\right)\frac{d}{dt}\left(V(x,t)+\frac{\mu}{en(x)}J(x,t)\right)\right|_{x=q_i(Q,D)}\nonumber\\
 &\qquad + \left.\mu \frac{d}{dt} \left(V(x,t)-\frac{(1-\mu)}{en(x)}J(x,t)\right)\right|_{x=q_e(Q,D)}\nonumber\\
 &= \left(1-\mu\right)\left(\frac{\partial V}{\partial t} + \frac{\partial q_i}{\partial t}\cdot\nabla V +
 \frac{\mu}{en}\frac{\partial}{\partial t}\left(\frac{J}{n}\right)\right. \label{eq:qddot}\\
 &\qquad \left.+\frac{\mu}{e}\frac{\partial q_i}{\partial t}\cdot\nabla \left(\frac{J}{n}\right) \right)
 +\mu \left(\frac{\partial V}{\partial t}+ \frac{\partial q_e}{\partial t}\cdot\nabla V\right.\nonumber\\
 &\qquad\left.  -  \frac{(1-\mu)}{en}\frac{\partial}{\partial t}\left(\frac{J}{n}\right)-\frac{(1-\mu)}{e}\frac{\partial q_e}{\partial t}\cdot\nabla \left(\frac{J}{n}\right) \right)\,.
\nonumber
\end{align}
From Eqs.~\eqref{eq:qiqefirstorder}, we can find explicit expressions for the time derivatives of the $q_s(Q,D)$, 
\begin{align}
\frac{\partial q_i}{\partial t} &= \dot{Q} + \frac{\mu}{en_0}\dot{D}\longrightarrow V + \frac{\mu}{en} J\label{eq:qideriv}\\
\frac{\partial q_e}{\partial t} &= \dot{Q} - \frac{1-\mu}{en_0}\dot{D}\longrightarrow V - \frac{1-\mu}{en} J
\,.
\label{eq:qederiv}
\end{align}
Inserting these expression into \eqeqref{qddot},  we find after some algebra that 
\begin{equation}
\ddot{Q}(a,t) \longrightarrow \frac{\partial V}{\partial t} + (V\cdot\nabla) V + \frac{\mu (1-\mu)}{n e^2}(J\cdot\nabla)\left(\frac{J}{n}\right)
\,.
\end{equation}

Next we Eulerianize the interaction terms of \eqeqref{qeqextmhd} using  \eqeqref{invLEmap}  (up to first order in $\mu$) and \eqeqref{qiqefirstorder}. The result is
\begin{align}
& \frac{ne}{c}\left[\left(V_j+\frac{\mu}{e n}J_j\right)
\frac{\partial A_j}{\partial x^k} - c\partial_k\phi - \frac{\partial A_k}{\partial t} -\frac{\partial q_i}{\partial t}\cdot\nabla A_k \right]\nonumber\\[.1in]
& + \frac{ne}{c} \left[\left(-V_j+\frac{(1-\mu)}{e n}J_j\right)
\frac{\partial A_j}{\partial x^k} + c\partial_k\phi + \frac{\partial A_k}{\partial t}\right.\nonumber\\
&\qquad\qquad \left. +\frac{\partial q_e}{\partial t}\cdot \nabla A_k \right]\,, 
\end{align}
which, after substitution using Eqs.~\eqref{eq:qideriv} and \eqref{eq:qederiv},  yields
\begin{equation}
 \frac{1}{c}\left(J_j \frac{\partial A_j}{\partial x^k} - J_j \frac{\partial A_k}{\partial x^j} \right) = \frac{\left(J\times \left(\nabla\times A\right)\right)_k}{c}\,.
\end{equation}

The full Eulerian version of the equation of motion for the velocity of \eqeqref{qeqextmhd}, also referred to as the {\it momentum equation} is  
\begin{eqnarray}
 n m \left(\frac{\partial V}{\partial t} + (V\cdot\nabla) V\right) &=& -\nabla p  + \frac{J\times B}{c}
\label{eq:momextmhd} 
\\
 && \hspace{.25 cm}  -
 \frac{m_e}{e^2}(J\cdot\nabla)\left(\frac{J}{n}\right)\,.
  \nonumber
\end{eqnarray}
Note, it was shown in Ref.~\onlinecite{kimura} that the last term of \eqeqref{momextmhd} is necessary for energy conservation.

Next, varying the action with respect to $D_k$ yields
\begin{align}
0 &= \frac{m_i \mu}{n_0 e^2} \ddot{D}_k(a,t)+ \frac{(1-\mu)}{en_0}\partial_k p_e - \frac{\mu}{en_0}\partial_k p_i
\nonumber\\[.1in]
& + \mu\left[\left(-\frac{1}{c} \frac{d}{dt}A_k(x,t) -\partial_k\phi(x,t) + \frac{1}{c}
\left(\dot{Q}_j(a,t)\right.\right.\right.\nonumber\\
& \qquad \left.\left.\left. \left.+\frac{\mu}{e n_0}\dot{D}_j(a,t)\right)
\frac{\partial A_j(x,t)}{\partial x^k}\right)\right]\right|_{x=q_i(Q,D)}\nonumber\\[.1in]
& + (1-\mu)\left[\left(-\frac{1}{c} \frac{d}{dt}A_k(x,t) -\partial_k\phi(x,t) +\frac{1}{c}
\left(\dot{Q}_j(a,t)\right.\right.\right.\nonumber\\
&\quad\left.\left.\left.\left. -\frac{(1-\mu)}{e n_0}\dot{D}_j(a,t)\right)
\frac{\partial A_j(x,t)}{\partial x^k}\right)\right]\right|_{x=q_e(Q,D)}\,. 
\label{eq:deqextmhd}
\end{align}
This time the Jacobians of the internal energies are varied using $\delta q_e=-(1-\mu)\delta D/(en_0)$ and $\delta q_i=\mu\delta D/(en_0)$, which again follow from  Eqs.~\eqref{eq:qiqefirstorder}.  Note, it is for this reason that only the electron pressure appears to leading order in Ohm's law for extended MHD. 

Eulerianizing the $\ddot{D}$ term in \eqeqref{deqextmhd} yields 
\begin{align}
&\frac{m_i\mu}{n_0 e^2}\ddot{D}(a,t) = \frac{m_i \mu}{e^2 n}\left(\frac{\partial J}{\partial t} + (J\cdot\nabla)V - (J\cdot\nabla)\left(\frac{J}{n}\right)\right)
\nonumber\\
& \qquad\qquad +
\frac{m_i\mu}{e^2}(V\cdot\nabla)\left(\frac{J}{n}\right) + \frac{m_i \mu}{e^2 n^2}J(V\cdot \nabla)n
\label{eq:Edeqextmhd} 
\end{align}
where we have used the continuity equation \eqeqref{newconteq} to eliminate the time derivative of $n$ and kept leading  order in $\mu$ terms.
The interaction terms in \eqeqref{deqextmhd} reduce to
\begin{equation}
E +\frac{V\times(\nabla\times A)}{c} - \frac{(1-2\mu)}{e n  c} J\times(\nabla\times A)\,.
\label{eq:Eohm}
\end{equation}
In Eqs.~\eqref{eq:Edeqextmhd} and \eqref{eq:Eohm}   
we see the presence of some terms involving $\mu$, in front of $\ddot{D}$ and $J \times B$,  respectively. However, in the latter case it occurs in the factor $(1-2\mu)$ and since our ordering is $\mu << 1$, we can drop the $\mu$-dependence in 
Eq.~\eqref{eq:Eohm}, to lowest order. However, in Eq.~\eqref{eq:Edeqextmhd}, we cannot throw out all the terms that depend on $\mu$ since the factor ${\mu m_i}/({n e^2})$ cannot be cast into a dimensionless form, and hence one cannot invoke the ordering $\mu << 1$ here. Post-variation, the discrepancy in the order of the derived terms, i.e. the existence of these anomalous terms, has also been observed elsewhere\cite{RGL83}.


The Eulerian version of the equation of motion of the current \eqeqref{deqextmhd} (after keeping zeroth order in $\mu$), also known as {\it generalized Ohm's law}, is then 
\begin{align}
E +\frac{V\times B}{c} & =
\frac{m_e}{e^2 n}\left(\frac{\partial J}{\partial t} + (J\cdot\nabla)V - (J\cdot\nabla)\left(\frac{J}{n}\right)\right)
\nonumber\\
&  +\frac{(J\times B)}{e n c} -\frac{\nabla p_e}{e n} +
\frac{m_e}{e^2}(V\cdot\nabla)\left(\frac{J}{n}\right)\nonumber\\
&\hspace{.5 cm}+\frac{m_e}{e^2 n^2}J(V\cdot\nabla)n\,.
\label{eq:eohm}
\end{align}
The last two terms on  the right hand side of \eqeqref{eohm} can be combined to give $\left(m_e/(e^2 n)\right)(V\cdot\nabla)J$ and since $\nabla\cdot J=0$, we can add a $V (\nabla \cdot J)$ term without changing the result, and combine most terms in the divergence of the tensor
$VJ+JV$ to obtain the following equation:
\begin{align}
E +\frac{V\times B}{c} &=
\frac{m_e}{e^2 n}\left(\frac{\partial J}{\partial t} + \nabla\cdot(VJ+JV)\right)\nonumber\\
&\hspace{-.2cm} -\frac{m_e}{e^2 n} (J\cdot\nabla)\left(\frac{J}{n}\right)+\frac{(J\times B)}{e n c} -\frac{\nabla p_e}{e n}\,.
\label{eq:ohmextmhd}
\end{align}
Equations \eqref{eq:momextmhd} and \eqref{eq:ohmextmhd} constitute the extended MHD model.

\subsection{Hall MHD}

Hall MHD is a limiting case, for which previous work of an  action functional nature exists\cite{06hirota,13yoshida}.  Here we obtain the actional functional by expanding and retaining only terms up to zeroth order in $\mu$, i.e., if we neglect the electron inertia $(m_e\rightarrow 0)$, the action  of \eqeqref{newgenact} reduces to
\begin{align}
 S & =  -\frac{1}{8\pi} \int dt \int d^3x\; \left|\nabla\times A(x,t)\right|^2\nonumber\\
 & +\int dt \int d^3a\; n_0\bigg[\frac{1}{2} m |\dot{Q}|^2(a,t) 
 \nonumber\\
 &\hspace{1.5cm} - \mathfrak{U}_{i}\left(\frac{n_0}{\mathcal{J}_i(Q)},s_{i0}\right)
-  \mathfrak{U}_{e}\left(\frac{n_0}{\mathcal{J}_e(Q,D)},s_{e0}\right)\bigg]\nonumber\\
& + \int dt \int d^3x \int d^3a\; n_0 \left\{\delta\!\left(x-Q(a,t)-\frac{1}{e n_0}D(a,t)\right)\right.\nonumber\\
&\hspace{.2cm} \times\left.  \left[-\frac{e}{c}\left(\dot{Q}(a,t)-\frac{1}{e n_0}\dot{D}(a,t)\right)\cdot
                                    A(x,t)+e\phi(x,t)\right]\right\}
                                    \nonumber\\
& + \int dt \int d^3x \int d^3a\; n_0 \bigg\{\delta\left(x-Q(a,t)\right) \nonumber\\
&\hspace{1.5cm} \times \left[\frac{e}{c}\dot{Q}(a,t)\cdot A(x,t)-e\phi(x,t)\right]\bigg\}\,.
\end{align}
and Eqs.~\eqref{eq:qiqefirstorder} become
\begin{align}
 q_i(Q,D) &= Q(a,t) \nonumber\\
 q_e(Q,D) &= Q(a,t) - D(a,t)/(en_0)
 \label{eq:qiqehall}
\end{align}
Observe we have also replaces $m_i$ by $m$ in the kinetic energy term, which is correct to leading order in $\mu$. 

 The inverse maps required for Eulerianizing the equations of motion reduce to
\begin{align}
 \dot{Q}(a,t) &= V(x,t)\biggl{|}_{x=q_i=Q}
 \nonumber \\
\dot{D}(a,t)&= en_0 V(x,t)\biggl{|}_{x=q_i=Q}\\
&\qquad - \left.en_0\left(V(x',t)-\frac{J(x',t)}{en(x',t)}\right)\right|_{x'=q_e(Q,D)}\, .
\nonumber
\end{align}
Following the procedure outlined in the previous section for extended MHD, we arrive at what is commonly referred to as  Hall MHD, 
\begin{eqnarray}
 &&n m \left(\frac{\partial V}{\partial t} + (V\cdot\nabla) V\right) = -\nabla p + \frac{J\times B}{c}
\\ 
&&E +\frac{V\times B}{c} = \frac{J\times B}{n e c}- \frac1{ne}\nabla p_e\,,
\end{eqnarray}
which are the usual forms of the momentum equation and Ohm's law for Hall MHD. 

\subsection{Electron MHD}
\label{ssec:electronMHD}

Electron MHD \cite{Lighthill,braginskii65,rudakov,ilgisonis1999} is another limiting case where we neglect the ion motion completely. This theory is used to describe the short time scale motion of the electrons in a neutralizing ion background. Since the ions are immobile, we require $\dot{q}_i = 0$ and $q_i = q_i(a)$.  Also, we require that there be no electric field and, consequently,  we neglect $\phi$ from the action. In this case, using the $Q$, $D$ formulation of the previous sections is redundant since there is only a single fluid. From $\dot{q}_i = 0$ we find  $\dot{D} = -\left(e n_0 m/m_e\right) \dot{Q}$. (The same relation holds between $Q$ and $D$ up to an additive constant which represents the constant position of the ion). In addition, the Lagrange-Euler map takes on the simple form
\begin{equation}
\label{eq:inertialQI}
v_e(x,t) =\left.\left(1+\frac{1}{\mu}\right)\dot{Q}(a,t)\right|_{a = q^{-1}_e(x,t)}\,,
\end{equation}
 where
 \begin{equation}
 \label{eq:inertialQII}
 q_e(a,t) = (1+\frac{1}{\mu})Q(a,t)\,.
 \end{equation}
The remaining terms in the action are
\begin{align}
 S & =  -\frac{1}{8\pi} \int dt \int d^3x\; |\nabla\times A(x,t)|^2\nonumber\\
 & \hspace{-.2in} +\int dt \int d^3a\; n_0\left[\frac{1}{2} m_e |\dot{q}_e|^2(a,t) - \mathfrak{U}_{e}\left(\frac{n_0}{\mathcal{J}_e(Q)}\right)\right]
 \nonumber\\
& - \int dt\int d^3x \int d^3a \; \delta \left(x-q_e\right) \frac{e n_0}{c} \dot{q}_e\cdot A(x,t)\,, 
\end{align}
which is essentially the same  action as that of \rcite{ilgisonis1999}. It is also straight-forward to express this  action in terms of $Q$ using Eqs.~\eqref{eq:inertialQI} and \eqref{eq:inertialQII}.

Upon varying the action (either in terms of $q_e$ or $Q$) and Eulerianizing the following equation of motion and constraint are obtained:  
\begin{align}
 m_e\left(\frac{\partial v_e}{\partial t} + v_e\cdot\nabla v_e\right) + \frac{e}{c}\frac{\partial A}{\partial t} =& \frac{e}{c} (v_e\times B) -\frac{\nabla p_e}{n}
\nonumber\\
\nabla\times B =& -\frac{4 \pi}{c} e n v_e
\,,
 \nonumber
\end{align}
which are the usual equations of electron MHD. 

\section{Noether's theorem}
\label{sec:noether}

In this section we will investigate  the invariants of the action functionals for the quasineutral L\"{u}st equations of \eqeqref{newgenact} and the extended MHD system of \eqeqref{actionextmhd} using Noether's theorem.  Note that both actions can be expressed either in terms of $\left(Q,D\right)$ or in terms of
$\left(q_{i},q_{e}\right)$, which are related through a simple linear transformation, e.g., \eqeqref{newvarhom}. Furthermore, both sets of variables obey the Eulerian Closure Principle. Hence, it is equivalent to work with an action expressed in terms of either set of variables. For convenience,  we shall work with the latter set, as the Euler-Lagrange maps are easier to apply.
Noether's theorem states that if an action is invariant under the transformations
\begin{equation}
q'_s=q_s+K_s\left(q_s,t\right);\quad t'=t+\tau\left(t\right),
\end{equation}
i.e.,
\begin{align*}
S &=\int_{t_{1}}^{t_{2}} dt \int d^3z\; \mathcal{L}\left(q_s,\dot{q}_{s},z,t\right)\\
&=\int_{t'_{1}}^{t'_{2}} dt' \int d^3z'\; \mathcal{L}\left(q'_s,\dot{ q}'_s,z',t'\right)\,,
\end{align*}
then there exist constants of motion given by
\begin{equation}
C=\int d^3z\left[\tau\left(\frac{\partial\mathcal{L}}{\partial{\dot{q}}_{s}}\cdot\dot{{q}}_{s}-\mathcal{L}\right) 
-{ K}_{s}\cdot\frac{\partial\mathcal{L}}{\partial\dot{{q}}_{s}}\right]\,,
\end{equation}
where the index $s$ represents the number of independent variables $q$ in the system. Our actions are mixed Lagrangian and Eulerian, so the variable $z$ can denote $a$ or $x$.\\

\subsection*{1. Time translation}

It is straight-forward to see that the action is invariant under time translation
with
\[
K_s=0;\quad\tau=1 \,.
\]
The corresponding constant of motion, the energy, is found to be
\begin{align*}
\mathcal{E}= & \int d^3x\left[\frac{\left|\nabla\times{A}\right|^{2}}{8\pi}+ \sum_{s}\int d^3a\left(\frac12 n_{0}m_{s}{|\dot{q}_s|^{2}}\right. \right.\\
&  \left.\left.\qquad\qquad +n_{0} \,\mathfrak{U}_{s}\left(\frac{n_{0}}{\mathcal{J}_{s}},s_{s0}\right)\right)  \right] \, .
\end{align*}
Using suitable Lagrange-Euler maps to express our answer in terms of the Eulerian variables $\left\{ n,V,J\right\} $,
we obtain
\begin{equation}
\mathcal{E}=\int d^3x \left[\frac{|B|^{2}}{8\pi}+n\mathfrak{U}_{i}+n\mathfrak{U}_{e}+mn\frac{|V|^{2}}{2}+\frac{m_{e}m_i}{mne^{2}}\frac{|J|^{2}}{2}\right]
\end{equation}
for the quasineutral L\"{u}st model and
\begin{equation}
\mathcal{E}=\int d^3x \left[\frac{|B|^{2}}{8\pi}+n\mathfrak{U}_{i}+n\mathfrak{U}_{e}+mn\frac{|V|^{2}}{2}+\frac{m_{e}}{ne^{2}}\frac{|J|^{2}}{2}\right]
\end{equation}
for the extended MHD model. Note that the two energies are different since the extended MHD model
includes the mass ratio ordering.\\

\subsection*{2. Space translation}

Space translations correspond to
\[
K_s=k;\quad\tau=0\, ,
\]
where $k$ is an arbitrary constant vector. Under space translations,
the constant of motion is the momentum, which is found to be
\begin{align*}
P &=k\cdot\int d^3a \left(n_{0}m_{i}\dot{q}_i+n_0 m_e\dot{q}_{e}\right)\\
&+k\cdot\int d^3x\; \frac{e}{c}{A}\left\{ \int n_0\left[\delta\left(x- q_i\right)-\delta\left(x-q_e\right)\right]d^3a\right\}\, ,
\end{align*}
Using the Lagrange-Euler  maps one can show that
\[
P= k\cdot\int d^3x\; nm V
\]
is the conserved quantity. Note that $k$ is entirely arbitrary,
and hence we see that the total momentum
\begin{equation}
P=\int d^3x\;\rho V
\end{equation}
is conserved. This is also evident from the corresponding dynamical equation for $V$. \\

\subsection*{3. Rotations}

The actions are also invariant under rotations which correspond to
\[
K_s=k\times q_s;\quad\tau=0\, ,
\]
Following the same procedure as before, we have
\[
\mathfrak{L}=k\cdot\int d^3x\; nm\,  r\times V\, ,
\]
and since we know that $k$ is arbitrary, we conclude that the
angular momentum given by
\begin{equation}
\mathfrak{L}=\int d^3x\,  \rho\,  r\times V
\end{equation}
is a constant of motion.\\

\subsection*{4. Galilean boosts}

When discussing boosts, we have to consider that the action may
remain invariant even when the following holds
\begin{align*}
S &=\int_{t_{1}}^{t_{2}} dt \int d^3z\; \mathcal{L}\left(q_s,\dot{q}_{s},z,t\right)\\
&=\int_{t'_{1}}^{t'_{2}} dt' \int d^3z' \left(\mathcal{L}\left(q'_s,\dot{ q}'_s,z',t'\right)+\partial_{\mu}\lambda^{\mu}\right)\, ,
\end{align*}
because the second term vanishes identically. In all the previous
derivations of the constants of motion, the infinitesimal transformations
did not involve time explicitly.  A boost, though, corresponds to
\[
K_s=u t;\quad\tau=0\, ,
\]
where $u$ is an arbitrary constant velocity. For a Galilean boost in a one-fluid
model, the corresponding invariant quantity is given by
\[
\mathcal{B}=\int d^3a\; mn\left(q-\dot{q} t\right)\, ,
\]
and since we have two different species, this generalizes to
\[
\mathcal{B}=\sum_{s}\int d^3a\; m_s n_s \left(q_s-\dot{q}_s t\right)\, .
\]
Using the corresponding Lagrange-Euler maps, the Eulerianized expression is given by
\begin{equation}
\mathcal{B}= \int d^3x\; \rho\left(x-V t\right)\, .
\end{equation}

\section{Conclusions}
\label{sec:concl}

In this paper, we derived several fluid models from a general two-fluid action functional. All approximations, ordering schemes, and changes of variables were done in the action functional before Hamilton's principle was invoked. We defined a new set of Lagrangian variables, and under the assumption of quasineutrality, we constructed a new set of nonlocal Lagrange-Euler maps assuring that our  Lagrangian equations of motion can be  Eulerianized. Lastly, we derived several conservation laws for these models using Noether's theorem.


The novel nonlocal Lagrange-Euler map of this paper is of particular general importance.  Usual Lagrange-Euler  maps  (also known as momentum maps)  entail the advection of various quantities by a single velocity field and this can be traced to the  algebraic structure of the Poisson bracket written in terms of Eulerian variables (see e.g.\ Ref.~\onlinecite{morrison1998}).  For single fluid models like MHD the Poisson bracket\cite{morrison80} has semi-direct product structure, which occurs in a variety of fluid contexts (e.g.\ Refs.~\onlinecite{rosensteel,HK,MM84}).  However, many systems do not possess this semi-direct product structure  (e.g.\ Refs.~\onlinecite{HHM87,GCPP01,TMWG08,pjmTT13}) and indeed a general theory of algebraic extensions was given in Ref.~\onlinecite{pjmT00}.  It is the selection of the set of observables and the ECP that give rise to the general algebras underlying Poisson brackets.   Detailed construction of  general algebras of  Ref.~\onlinecite{pjmT00} will be reported in future publications, along with derivations of other single fluid models including gyroviscosity\cite{morrison2014hamiltonian,lingam14}. 

\begin{acknowledgments}
IKC, ML, PJM, and  RLW received support from the  U.S.\  Dept.\ of Energy Contract \# DE-FG05-80ET-53088.
AW thanks the Western New England University Research Fund for support.
\end{acknowledgments}

\end{document}